\documentclass[aps,twocolumn,amsmath,amssymb,showpacs]{revtex4-2}
\usepackage{graphicx}
\usepackage{xcolor}
\usepackage{siunitx}
\usepackage{hyperref}
\usepackage{braket}
\usepackage{booktabs}
\usepackage{multirow}

\begin{document}

\title{Cavity-enhanced spectroscopy of individual nuclear spins in a dense bath}
\author{Alexander Ulanowski}
\author{Olivier Kuijpers}
\author{Benjamin Merkel}
\author{Adrian Holz\"apfel}
\author{Andreas Reiserer}
\email{andreas.reiserer@tum.de}

\affiliation{Max-Planck-Institut f\"ur Quantenoptik, Quantum Networks Group, Hans-Kopfermann-Stra{\ss}e 1, D-85748 Garching, Germany}
\affiliation{Technical University of Munich, TUM School of Natural Sciences, Physics Department and Munich Center for Quantum Science and Technology (MCQST), James-Franck-Stra{\ss}e 1, D-85748 Garching, Germany}

\begin{abstract}
Echo-based spectroscopy of the superhyperfine interaction of an electronic spin with nuclear spins in its surroundings enables detailed insights into the microscopic magnetic environment of spins in solids. Still, it is an outstanding challenge to resolve individual nuclear spins in a dense bath, in which many of them exhibit a comparable coupling strength. This simultaneously requires a high spectral resolution and a large signal-to-noise ratio. However, when probing spin ensembles, dipolar interactions between the dopants can lead to a concentration-dependent trade-off between resolution and signal. Here, we fully eliminate this limitation of previous optical-echo-envelope-modulation spectroscopy experiments by integrating the emitters into a high-finesse resonator, which allows for strong optical echoes even at very low concentrations. To demonstrate its potential, the technique is applied to erbium dopants in yttrium-orthosilicate (Er:YSO). Achieving an unprecedented spectral resolution enables precise measurements of the superhyperfine interaction with four of the Y nuclear spins densely surrounding each emitter. The achieved boost of the signal, enabled by the resonator, allows for extending the approach to the lowest concentration possible --- to the level of single dopants, thereby providing a tool for detecting and studying individual nuclear spins. Thus, our technique paves the way for an improved understanding of dense nuclear spin baths in solids.
\end{abstract}

\maketitle

\section{Introduction} \label{sec:Intro}

Spins in solids enable many applications in quantum technology, including networking, information processing, and metrology~\cite{awschalom_quantum_2018}. Via their potential for on-chip integration, they show particular promise for the up-scaling of quantum devices~\cite{li_heterogeneous_2024}. To this end, it is paramount to understand their interaction with the host material~\cite{wolfowicz_quantum_2021}. To achieve low decoherence rates, one typically uses crystals with a very low concentration of paramagnetic impurities and magnetic-field-insensitive states \cite{zhong_optically_2015, ortu_simultaneous_2018} or conditions where the electronic spins are frozen to the ground state, i.e. a large magnetic field and ultralow temperature~\cite{rancic_coherence_2018, le_dantec_twenty-threemillisecond_2021}. In this situation, the dominant source of decoherence of spin-~\cite{wolfowicz_quantum_2021} and narrow-linewidth optical transitions~\cite{bottger_effects_2009, car_superhyperfine_2020, ulanowski_spectral_2022, mccullian_quantifying_2022} stems from the superhyperfine interaction, i.e. the dipolar coupling to nuclear spins in the host crystal surrounding the emitter. Isotopic purification can alleviate this decoherence mechanism, but only in a few materials~\cite{maurer_room-temperature_2012, saeedi_room-temperature_2013}. 

Instead of merely acting as a source of decoherence, however, nuclear spins of the host material can also give access to additional resources in the form of nuclear spin registers~\cite{dutt_quantum_2007,robledo_high-fidelity_2011,ruskuc_nuclear_2022} and robust quantum network memories \cite{reiserer_robust_2016} if techniques to reliably control them are implemented. This has been achieved via the superhyperfine interaction with a controlled electronic spin~\cite{wolfowicz_quantum_2021, maurer_room-temperature_2012, kornher_sensing_2020, uysal_coherent_2023, dutt_quantum_2007, robledo_high-fidelity_2011,reiserer_robust_2016} or hyperfine state \cite{zhong_quantum_2019, ruskuc_nuclear_2022}.

So far, however, the technique was restricted to systems in which the superhyperfine interaction of a single nuclear spin ~\cite{maurer_room-temperature_2012, kornher_sensing_2020, uysal_coherent_2023}, or a small number of them \cite{dutt_quantum_2007,robledo_high-fidelity_2011,ruskuc_nuclear_2022,reiserer_robust_2016}, differed significantly from that of the others and/or was comparably weak. This way, in a dilute bath where only a single spin had a coupling exceeding $\SI{70}{\kilo\hertz}$, up to 27 nuclear spins could be resolved using tailored dynamical decoupling sequences~\cite{abobeih_atomic-scale_2019}. However, these techniques have not been successfully applied to resolve the individual nuclear spins of a dense bath that exhibit comparable and strong superhyperfine interactions \cite{kornher_sensing_2020, uysal_coherent_2023}. As a specific example, in Er:YSO, where the coupling of 15 spins exceeds 70 kHz, only a fast decay of the coherence caused by the bath is observed in dynamical decoupling spectroscopy~\cite{uysal_coherent_2023}. This limitation of the technique has restricted the number of accessible nuclear spins and the host materials that can be investigated.

An alternative spectroscopic method has been introduced recently to study nuclear spins in Er:YSO~\cite{car_selective_2018}, which we term optical echo envelope modulation (OEEM) to reveal the similarity to the well-known electron spin echo envelope modulation (ESEEM) spectroscopy~\cite{mims_envelope_1972, guillot-noel_direct_2007, deligiannakis_electron_2000}. However, using pulses in the optical rather than the microwave domain has two key advantages: First, commercially available single-photon detectors facilitate the detection of tiny echo signals. Second, in contrast to microwave resonators for high-sensitivity ESEEM with their constrained bandwidth~\cite{billaud_microwave_2023}, optical excitation allows for a full tunability of the transition frequency and thus for complete flexibility of the amplitude and orientation of the magnetic bias field. This enables a high degree of selectivity, meaning that the modulation is caused only by nuclear spins located in a specific direction determined by the magnetic field~\cite{car_selective_2018}. The effect is particularly relevant in systems with a highly anisotropic magnetic interaction, as e.g. encountered with rare-earth dopants, allowing for optical spectroscopy of nuclear spins in a dense bath. However, previous experiments were only performed on large ensembles of emitters~\cite{car_selective_2018}, which limited both the spatial and spectral resolution.

Here, we overcome these limitations by coupling the emitters to a high-finesse optical resonator \cite{merkel_coherent_2020}. While this approach can be applied to many material systems \cite{reiserer_colloquium_2022}, specifically we study the superhyperfine interaction of yttrium nuclear spins surrounding erbium dopants in Er:YSO. Compared to previous works \cite{car_optical_2019}, resonator integration allows reducing the dopant concentration and thus eliminates decoherence from the interaction of the electronic spins within the ensemble that cannot be decoupled efficiently \cite{merkel_dynamical_2021}. In this way, our experiment characterizes the superhyperfine interaction between electronic and nuclear spins with an unprecedented spectral resolution. In addition, owing to the efficient detection of the photon echo enabled by the resonator, the signal-to-noise-ratio is improved dramatically, which enables measurements down to the level of single emitters. With this, we observe significantly modified superhyperfine spectra for dopants that belong to satellite peaks in which the site properties are changed because of the proximity of europium co-dopants~\cite{ulanowski_spectral_2024}. In conclusion, our work thus enables exploring the interaction of individual emitters with a dense bath of nuclear spins, giving insights into their microscopic environment as opposed to the ensemble-averaged couplings studied previously.

\section{Experimental setup}

Our experiments are based on erbium dopants in a \SI{10}{\micro\meter} thin membrane of YSO that is integrated into an optical resonator~\cite{ulanowski_spectral_2024}. Both the dopant and the host material are of particular interest for quantum applications, as erbium offers an optically addressable spin featuring an optical transition directly into the telecom C-band, where absorption in optical fibers is minimal~\cite{reiserer_colloquium_2022}. This transition may feature a particularly narrow linewidth~\cite{bottger_effects_2009} and may even reach its lifetime limit when coupled to a high-finesse resonator~\cite{merkel_coherent_2020, ulanowski_spectral_2022}. For this reason, experiments that resolve and control single Er emitters have become a recent focus of research \cite{dibos_atomic_2018, ulanowski_spectral_2022, xia_tunable_2022, yu_frequency_2023, ourari_indistinguishable_2023, gritsch_purcell_2023, deshmukh_detection_2023, ulanowski_spectral_2024}. 

YSO, on the other hand, is a well-established host material for rare-earth dopants in a quantum technology context~\cite{thiel_rare-earth-doped_2011}. The triply charged dopant ions may substitute a single yttrium ion with the same charge state and comparable ionic radius. Beyond that, YSO provides an environment with comparably low magnetic field fluctuations. With oxygen and silicon only featuring isotopes with non-zero spin at low abundances of $<1\%$ and $<5\%$, the main contribution to magnetic noise is $\mathrm{^{89}Y}$ with natural abundance of unity. While it features a particularly small magnetic moment of about $14\%$ of the nuclear magneton, interaction with the yttrium bath may still constitute the major limitation to the achieved coherence times~\cite{holzapfel_optical_2020,nicolas_coherent_2023}, even at magnetically insensitive transitions that hold the current record of more than ten hours in qubit coherence~\cite{wang_nuclear_2025}.

\begin{figure}
    \centering
    \includegraphics{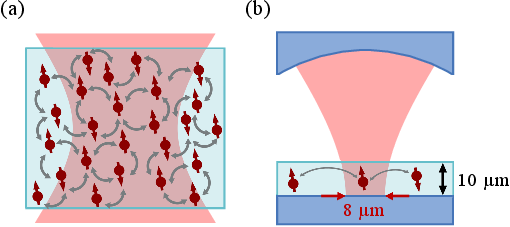}
    \caption{(a) Measuring photon echoes to study the superhyperfine interaction in a bulk sample requires a large volume and a high dopant concentration, which impedes the spatial and spectral resolution. (b) These limitations are overcome by integrating the emitters into an optical resonator, as the Purcell effect can enhance the emission such that individual spins can be resolved in a thin membrane at low dopant concentration.}
    \label{FIG:setup}
\end{figure}

The energy levels of erbium in YSO are organized in Kramers doublets with effective spin $1/2$, whose Zeeman interaction is anisotropic and thus depends on the orientation of an externally applied magnetic field. The magnetic moment may exceed the Bohr magneton by more than an order of magnitude; this makes erbium dopants a sensitive probe of their magnetic environment but also entails strong interactions between erbium spins that may severely limit the coherence~\cite{merkel_dynamical_2021} and thus the spectral resolution.

As illustrated in Figure~\ref{FIG:setup}, this limitation is overcome in our experiment by placing the sample into a high-finesse cavity, which allows recording optical echo signals down to the single-emitter level~\cite{ulanowski_spectral_2022}. Thus, the concentration of emitters can be kept low enough to observe coherence times of several hundred microseconds even at low magnetic fields~\cite{ulanowski_spectral_2024}, strongly exceeding those in comparable experiments with bulk crystals~\cite{car_selective_2018, car_superhyperfine_2020}. This enables optical detection of the superhyperfine interaction with an unprecedented spectral resolution of a few kilohertz. Beyond that, the spatial volume that is interrogated in such a cavity-coupled setup may be many orders of magnitude smaller than in bulk experiments. This spatial selectivity allows us to spectrally isolate individual emitters within the inhomogeneous line, such that probing of individual dopants becomes possible.

Our sample consists of YSO with trace amounts of erbium at an estimated concentration $<\mathrm{1~ppm}$. It is co-doped with 100~ppm europium to engineer the inhomogeneous line to facilitate the spectral resolution and control of hundreds of individual emitters~\cite{ulanowski_spectral_2024}. A membrane is obtained via polishing the bulk crystal to a thickness of \SI{10}{\micro\meter} along the crystal b-axis. The membrane is placed in a high-finesse Fabry-Perot resonator (linewidth \SI{65}{\mega\hertz} FWHM), which enables a Purcell-enhancement of $P \lesssim 100$~\cite{ulanowski_spectral_2024}. At the same time, our approach avoids the detrimental effects on optical coherence associated with proximity to surfaces that are encountered with nanostructured resonators~\cite{dibos_atomic_2018, xia_tunable_2022, yu_frequency_2023, ourari_indistinguishable_2023, gritsch_purcell_2023} or nanoparticles\cite{deshmukh_detection_2023}. Further details of the experimental setup are described in~\cite{ulanowski_spectral_2022, ulanowski_spectral_2024}.

In YSO, erbium is integrated in two crystallographic sites. During the experiment, site 1~\cite{bottger_effects_2009} is addressed resonantly at a wavelength of $\lambda\approx\SI{1536.48}{\nano\meter}$. This crystallographic site features two magnetically inequivalent classes of emitters that are related by a two-fold rotation around the b-axis of the crystal. As such, they will behave magnetically equivalent when the external magnetic field is parallel or orthogonal to this axis. The geometry of the superconducting solenoid used in our setup to apply the bias field currently restricts its direction to the b-axis.

\section{Superhyperfine spectroscopy by OEEM}\label{SEC:model}

To characterize the superhyperfine coupling of the electronic spin to nuclei surrounding the emitters, we rely on OEEM. The technique is similar to that studied in nuclear magnetic resonance~\cite{mims_envelope_1972} and ESEEM~\cite{guillot-noel_direct_2007, dikanov_electron_2024}. Specifically, when implementing an echo-sequence on a two-level system with an additional substructure, one observes interference effects resulting in a characteristic envelope-modulation of the temporal decay of the echo. The Fourier components and visibility of the envelope modulation depend on the splitting and branching ratio of the transitions of the substructure.

\begin{figure*}
    \centering
    \includegraphics{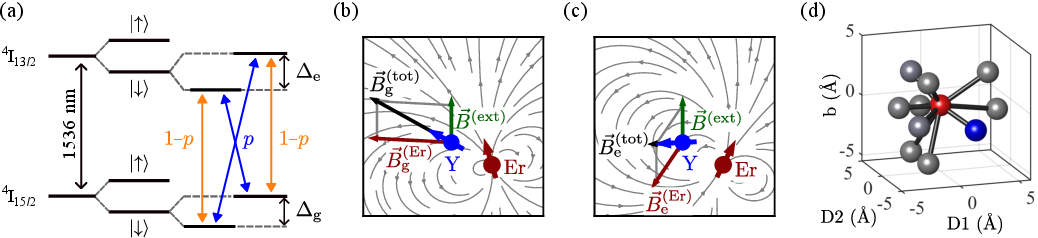}
    \caption{(a)~Level structure probed in the experiment. In an external magnetic field, the superhyperfine coupling leads to an additional substructure with detunings $\Delta_g$ and $\Delta_e$ for all Zeeman states $\ket{\uparrow}$ and $\ket{\downarrow}$, as sketched for the lower branches of the lowest $^4\text{I}_{13/2}$ and $^4\text{I}_{15/2}$ crystal field levels. The transition strengths depend on the branching fraction $p$. (b, c) Illustration of the coupling mechanism of Er emitters to individual Y spins. The optical excitation of an Er dopant from the lower spin state of the CF ground state, $\ket{^4\text{I}_{15/2},\downarrow}$ (b) to the lower spin state of the excited CF level $\ket{^4\text{I}_{13/2},\downarrow}$ (c) changes its magnetic moment. This rotates the magnetic field $\vec{B}^{\text{(tot)}}$ experienced by neighboring Y spins, which is the sum of the field generated by the Er dopant $\vec{B}^{\text{(Er)}}$ and the external field $\vec{B}^{\text{(ext)}}$. The resulting rotation of the quantization axis generates coherent superpositions of spin states in the yttrium bath. This leads to a characteristic envelope modulation when the optical coherence of the emitter is probed via an echo sequence. (d)~Erbium dopant (red, center) and its nearest yttrium neighbors (grey) in a coordinate system spanned by the D1 and D2 polarization extinction axes and the b-axis of the YSO crystal \cite{bottger_effects_2009}. While all Y spins have the same magnetic moment, only a very small subset contributes to the OEEM signal for a given field geometry. For a bias field of \SI{175}{\milli\tesla} along the b-axis of the crystal, for instance, essentially only a single Y spin (blue) will contribute. Corresponding numerical values are shown in Table~\ref{TAB:yttrium}.}
    \label{FIG:semiclassical}
\end{figure*}

This modulation effect can be observed using a Hahn-echo sequence. Here, a coherence is generated in an inhomogeneously broadened ensemble by applying a $\pi/2$-pulse at $t=0$ followed by a $\pi$-pulse at $t=\tau$. This will lead to a full rephasing and, thus, the emission of an echo at $t=2\tau$. In case the emitters behave as two-level systems, the echo decays in proportion to their coherence time. In contrast, when the emitters exhibit a sub-structure whose spread is smaller than the bandwidth of the pulses, the rephasing can be inhibited depending on $\tau$, leading to a characteristic envelope modulation $\Theta(\tau)$. While the technique is independent of the exact origin of the substructure, in this work, we focus on the superhyperfine interaction by which the optically addressed erbium states are magnetically coupled to a dense bath of $S=1/2$ yttrium nuclear spins. This leads to a splitting of the Er energy levels, as illustrated in Figure~\ref{FIG:semiclassical}~(a) for a single coupled Y nuclear spin. For a dense bath, each of the Er levels would split further into a manifold of sublevels (not shown).

An analytical formula for the envelope modulation $\Theta(\tau)$ of photon echoes in free-space experiments without a cavity has been derived in~\cite{car_superhyperfine_2020}:

\begin{equation}\label{EQU:modulation}
\begin{split}
    &\Theta(\tau)=\\
    &\prod_i \left\{ 1-\frac{\rho_i}{2}\left[1-\cos\left(2\pi\Delta_\mathrm{g}^{(i)}\tau\right)\right]\left[1-\cos\left(2\pi\Delta_\mathrm{e}^{(i)}\tau\right)\right] \right\}
\end{split}
\end{equation}

Here, the index $i$ runs over the nearest-neighbor Y spins, with $\Delta_\mathrm{g/e}^{(i)}$ being the superhyperfine splitting induced by the $i$-th Y nucleus when the Er dopant is in its ground state $\mathrm{g}$ or excited state $\mathrm{e}$, respectively. The quantity $\rho_i$ represents the branching contrast, which is defined in terms of the branching fraction $p_i$, i.e., the fraction of optical decays that induce a nuclear-spin flip of the $i-$th coupled Y nucleus:
\begin{equation}
    \rho_i= 4 p_i (1-p_i)
\end{equation}
For optical photon echos, the expression from Equation~\ref{EQU:modulation} needs to be squared to determine the echo intensity, which leads to a large number of frequency components. Alternatively, one can use the recently proposed fluorescence-detected photon echo technique to measure the corresponding field quantity and reduce the complexity of the spectrum, while at the same time increasing the signal-to-noise ratio for a small number of emitters~\cite{montjovet-basset_incoherent_2024}.

The resolution of the described OEEM spectroscopy via photon-echo measurements is directly tied to the optical coherence time $T_2$ of the emitters. In general, the amplitude of the echo can be approximated as a product of the envelope modulation $\Theta(\tau)$ with a stretched exponential
\begin{equation}
    A(\tau)=\Theta(\tau)\cdot\exp\left[-\left(\frac{2\tau}{T_2}\right)^\gamma\right],~~\gamma\geq1.
\end{equation}
As the superhyperfine splittings $\Delta_\mathrm{g/e}^{(i)}$ are estimated by taking the Fourier transform of this quantity, they can be determined with an accuracy $\Delta \nu \propto 1/T_2$. For emitters with a strong magnetic moment, which are favorable for sensing applications, a significant restriction on $T_2$ can arise from dopant-dopant-interactions, which scale inversely with the concentration of resonantly addressed emitters $n_\mathrm{eff}$~\cite{merkel_dynamical_2021}. In this regime, the accuracy of optical superhyperfine spectroscopy thus scales as $\Delta \nu \propto n_\mathrm{eff}$. In our work, we fully eliminate this limitation by coupling the emitters to a high-finesse cavity, such that Hahn-echo measurements can be performed at the level of few or even single emitters. Thus, the spectral resolution of our experimental technique outperforms earlier experiments with a large number of emitters~\cite{car_superhyperfine_2020} and enables measurements of the superhyperfine interaction with increased precision.

\section{Y nuclear spins around Er dopants in YSO}
After describing the OEEM technique, we now turn to the expected measurement outcomes in the studied material platform, in which each erbium dopants is surrounded by a dense bath of spin-$1/2$ yttrium nuclei. If the magnetic interaction and the relative position of all spins are known, $p_i$ and $\Delta_\mathrm{g/e}^{(i)}$ can be derived from first principles in an external magnetic bias field~\cite{car_selective_2018}. If the Zeeman interaction of erbium with this field is much stronger than the Er-Y interaction, one finds -- assuming purely dipolar magnetic interactions:
\begin{subequations}\label{EQU:carmodel}
\begin{align}
\label{EQU:Deltas}
\Delta_\mathrm{g/e}&=\frac{\mu_\mathrm{N}}{h}\left|g_\mathrm{Y}\vec{B}_\mathrm{g/e}^\mathrm{(tot)}\right|\\
\rho&=1-\left(\frac{ \vec{B}_\mathrm{g}^\mathrm{(tot)}\cdot\vec{B}_\mathrm{e}^\mathrm{(tot)}}{|\vec{B}_\mathrm{g}^\mathrm{(tot)}||\vec{B}_\mathrm{e}^\mathrm{(tot)}|}\right)^2\label{EQU:contrast}
\end{align}
\end{subequations}

Here, $\mu_\mathrm{N}$ is the nuclear magneton, $h$ is the Planck constant and $g_\mathrm{Y}\approx-0.2737(1)$ is the nuclear $g$-factor of $^{89}\mathrm{Y}$ in YSO ~\cite{becerro_revisiting_2004, hasler_89yttrium_1977}.
$\vec{B}_\mathrm{g/e}^\mathrm{(tot)}$ 
are the sum of the externally applied field $\vec{B}^\mathrm{(ext)}$ and the classically approximated dipolar field of erbium $\vec{B}^\mathrm{(Er)}_\mathrm{g/e}$ at the position $\vec{r}$ of the yttrium spin when the erbium ion is in its optical ground state or optical excited state, respectively, with

\begin{equation}
    \vec{B}^\mathrm{(Er)}_\mathrm{g/e}(\vec{r}) = - \frac{\mu_0}{4\pi} 
    \left[ \frac{\langle\vec{\mu}^\mathrm{(Er)}_\mathrm{g/e}\rangle}{r^3} - 3\frac{\left( \langle\vec{\mu}^\mathrm{(Er)}_\mathrm{g/e}\rangle \cdot \vec{r}\right)\vec{r}}{r^5} 
    \right].
\end{equation}
In this equation, $\mu_0$ is the vacuum permeability and $\langle\vec{\mu}^\mathrm{(Er)}_\mathrm{g/e}\rangle$ is the expectation value of the erbium electronic dipole moment which depends on the $g$-tensor $\mathbf{g}_\mathrm{g/e}$ of the respective Kramers doublet~\cite{sun_magnetic_2008} and the spin vector $\vec{S}_\mathrm{g/e}$:
\begin{equation}
    \langle\vec{\mu}^\mathrm{(Er)}_\mathrm{g/e}\rangle=-\mu_\mathrm{B}\cdot\mathbf{g}_\mathrm{g/e}\cdot\langle\vec{S}_\mathrm{g/e}\rangle.
\end{equation}
The expectation value of the spin vector $\langle\vec{S}_\mathrm{g/e}\rangle$ is derived from the spin Hamiltonian which is given by 
\begin{equation}
    \mathcal{H}_\mathrm{g/e}=\mu_\mathrm{B}\cdot\vec{B}^\mathrm{(ext)}\cdot\mathbf{g}_\mathrm{g/e}\cdot\vec{S}_\mathrm{g/e}
\end{equation}
with $\mu_\mathrm{B}$ being the Bohr magneton.

In the spin Hamiltonian, hence also in Equation~\ref{EQU:carmodel}, two implicit assumptions are made: First, the quantization axis of the erbium spin is fully defined by $\vec{B}^\mathrm{(ext)}$, which will hold for fields exceeding a few milli-Tesla. Second, yttrium-yttrium interactions are neglected, which is justified as relative corrections are expected to be of the order of $\mu_\mathrm{N}/\mu_\mathrm{B}\approx10^{-3}$. A detailed analysis of the assumptions in this model and its relationship to other models is given in~\cite{pignol_decoherence_2024}.

The expression~\ref{EQU:contrast} can be intuitively understood as a measure of how much the change of the quantization axis caused by the excitation of the erbium mixes the yttrium eigenstates. For (anti-)collinear fields $\vec{B}_\mathrm{g}^\mathrm{(tot)}$ and $\vec{B}_\mathrm{e}^\mathrm{(tot)}$, the quantization axis and therefore eigenstates do not change, corresponding to a branching contrast of $\rho=0$. The situation is different if the fields are non-collinear, as illustrated in Figure~\ref{FIG:semiclassical}~(b,c). In particular, when $\vec{B}_\mathrm{g}^\mathrm{(tot)}$ and $\vec{B}_\mathrm{e}^\mathrm{(tot)}$ are orthogonal, the new eigenbasis is mutually unbiased to the old one. Consequently, the new eigenstates are an equal-weight superposition of the old ones. This corresponds to a branching contrast of $\rho=1$ and, thus, the strongest echo envelope modulation. 

\begin{table}
\centering
\begin{tabular}{@{}c|S[table-format=1.2]c@{}|SSSc@{}|S[table-format=2.3]|S[table-format=3.0]S[table-format=3.0]c@{}|S@{}}
\toprule
	Spin&\multicolumn{1}{c}{\begin{tabular}[c]{@{}c@{}}$d$\\(\AA)\end{tabular}}
	&&\multicolumn{1}{c}{\begin{tabular}[c]{@{}c@{}}$\mathrm{D1}$\\(\AA)\end{tabular}}
	&\multicolumn{1}{c}{\begin{tabular}[c]{@{}c@{}}$\mathrm{D2}$\\(\AA)\end{tabular}}
	&\multicolumn{1}{c}{\begin{tabular}[c]{@{}c@{}}$\mathrm{b}$\\(\AA)\end{tabular}}
	&&$\bar{\rho}$&\multicolumn{1}{c}{\begin{tabular}[c]{@{}c@{}}$\left|\mathrm{A}_\mathrm{g}\right|$\\(\si{\kilo\hertz})\end{tabular}}
	&\multicolumn{1}{c}{\begin{tabular}[c]{@{}c@{}}$\left|\mathrm{A}_\mathrm{e}\right|$\\(\si{\kilo\hertz})\end{tabular}}
	&&$\rho_\mathrm{max}$\\
	\midrule
	$\mathrm{Y_{1}}$ & 3.40 &  & -0.65 & 3.23 & -0.81 &  & 0.00 & 660 & 530 &  & 0.02 \\
 	$\mathrm{Y_{2}}$ & 3.46 &  & -3.45 & 0.29 & 0.00 &  & 0.01 & 430 & 370 &  & 0.04 \\
 	$\mathrm{Y_{3}}$ & 3.51 &  & -1.67 & -1.87 & 2.45 &  & 0.01 & 480 & 440 &  & 0.03 \\
 	$\mathrm{Y_{4}}$ & 3.62 &  & 2.26 & -2.25 & -1.72 &  & 0.97 & 360 & 280 &  & 1.00 \\
 	$\mathrm{Y_{5}}$ & 3.72 &  & -1.78 & 2.16 & 2.45 &  & 0.07 & 290 & 240 &  & 0.51 \\
 	$\mathrm{Y_{6}}$ & 4.15 &  & -2.80 & -2.95 & -0.81 &  & 0.07 & 200 & 180 &  & 0.20 \\
 	$\mathrm{Y_{7}}$ & 4.70 &  & 3.93 & -0.38 & 2.55 &  & 0.00 & 210 & 200 &  & 0.02 \\
 	$\mathrm{Y_{8}}$ & 4.95 &  & -1.67 & -1.87 & -4.27 &  & 0.00 & 140 & 150 &  & 0.08 \\
 	$\mathrm{Y_{9}}$ & 5.10 &  & -1.78 & 2.16 & -4.27 &  & 0.00 & 200 & 190 &  & 0.02 \\
 	$\mathrm{Y_{10}}$ & 5.19 &  & 5.06 & 0.70 & -0.91 &  & 0.01 & 110 & 90 &  & 0.06 \\
 	$\mathrm{Y_{11}}$ & 5.46 &  & -1.02 & -5.10 & 1.64 &  & 0.00 & 150 & 120 &  & 0.03 \\
 	$\mathrm{Y_{12}}$ & 5.46 &  & 1.02 & 5.10 & 1.64 &  & 0.02 & 110 & 80 &  & 0.38 \\
 	$\mathrm{Y_{13}}$ & 5.50 &  & 3.28 & 2.86 & -3.36 &  & 0.00 & 110 & 100 &  & 0.03 \\
 	$\mathrm{Y_{14}}$ & 5.50 &  & 3.28 & 2.86 & 3.36 &  & 0.00 & 90 & 90 &  & 0.10 \\
 	$\mathrm{Y_{15}}$ & 5.74 &  & 3.93 & -0.38 & -4.17 &  & 0.00 & 80 & 70 &  & 0.10\\ \bottomrule
\end{tabular}
\caption{Nearest yttrium neighbors of erbium dopants in class I of site 1. Their relative position is derived from crystallographic data~\cite{maksimov_crystal_1970} and is expressed in the D1-D2-b coordinate system of YSO~\cite{bottger_effects_2009}. The numbering follows the distance $d$ from the emitter. At a bias field of \SI{175}{\milli\tesla} applied along the b-axis, only the $\text{Y}_4$ nuclear spin exhibits a large branching contrast $\bar{\rho}$ on the lower electronic spin transition $\ket{^4\text{I}_{15/2},\downarrow}\leftrightarrow\ket{^4\text{I}_{13/2},\downarrow}$. Notably, its coupling to the erbium spin when the erbium electron spin is in its ground state ($|\mathrm{A_g}|$) and excited state ($|\mathrm{A_e}|$), respectively, does not differ significantly from the other nearby yttrium spins. Lastly, we display the maximum branching contrast $\rho_\mathrm{max}$ reached for each spin when the external bias field is restricted to $\vec{B}^\mathrm{(ext)}\parallel\mathrm{b}$.}
\label{TAB:yttrium}
\end{table}

As the field generated by the Er dopant strongly depends on the position, the above condition of orthogonal fields cannot be obtained for all surrounding Y spins simultaneously. Instead, for many fields only a single Y will contribute to the modulation, even in a dense spin bath with several nuclei at comparable distances to the emitter. This situation is shown in Figure~\ref{FIG:semiclassical}~(d), which includes the ten closest Y neighbors of an Er dopant in site 1 derived from crystallographic data~\cite{maksimov_crystal_1970}. Their precise relative position and distance from the emitter are summarized in Table~\ref{TAB:yttrium}.

To exemplify that at specific fields, only single Y spins contribute to the OEEM signal, Table~\ref{TAB:yttrium} further includes the branching contrast $\bar{\rho}$ when a bias field of $B=\SI{175}{\milli\tesla}$ is applied along the b-axis of the crystal. In this situation, only one Y exhibits $\rho>0.07$. When the field magnitude is varied, the contribution of other spins gets more significant. Still, when the field direction is kept along b, the maximum branching contrast $\rho_\mathrm{max}$ stays very low for most spins and is most considerable for $\text{Y}_4$, $\text{Y}_5$ and $\text{Y}_{12}$. Thus, these three spins are expected to dominate the OEEM signal and can be clearly distinguished from the rest of the bath, even though the magnitude of their superhyperfine coupling $\left| A_\mathrm{g/e} \right|$ does not differ strongly from that of the other spins.
We note that yttrium neighbors at larger distances, which are not listed in Table \ref{TAB:yttrium}, can also reach a significant branching contrast. However, this requires a magnetic field close to zero. In this regime, the large number of contributing yttrium spins leads to a superhyperfine-induced collapse of the echo~\cite{car_superhyperfine_2020} that precludes measurements using our scheme.

To also detect other nuclear spins, one can optimize the magnetic field direction and amplitude. To quantify how well individual Y spins can be isolated, we define the spin prominence $\lambda_i$ of the $i$-th neighbor as:

\begin{equation}   \label{EQU:spin_prominence}
    \lambda_i=\frac{\rho_i}{\sum_{j\neq i}\rho_j}.
\end{equation}

\begin{figure}
    \centering
    \includegraphics[width=1\columnwidth]{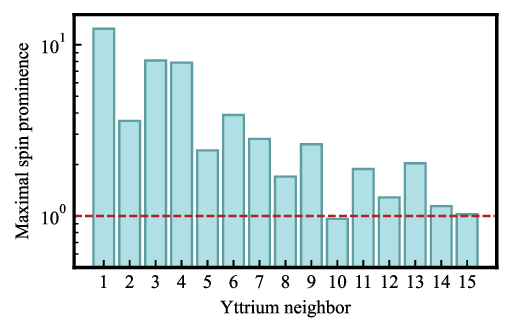}
    \caption{Maximum spin prominence of the different yttrium neighbors of erbium in site 1, as defined in Equation~\ref{EQU:spin_prominence}, when the magnitude and direction of the external magnetic bias field can be chosen freely. The yttrium neighbors are ordered by increasing distance to the erbium spin, see also table \ref{TAB:yttrium}.}
    \label{FIG:spin_prominence}
\end{figure}

Thus, $\lambda_i$ measures how much stronger $\text{Y}_i$ contributes to the OEEM signal as compared to all other Y spins. We now maximize $\lambda_i$ over all possible bias field directions and magnitudes; the result is shown in Figure~\ref{FIG:spin_prominence}. For the majority of the fifteen closest Y neighbors, a prominence of at least unity can be reached. This means that the modulation signal of the spin of interest dominates over the contributions of all other spins combined, which entails that OEEM spectroscopy enables highly selective spectroscopic investigations of individual nuclear spins even when they belong to a dense bath.

\section{OEEM spectroscopy with small ensembles of erbium dopants}
\begin{figure*}
    \centering
    \includegraphics[width=1\textwidth]{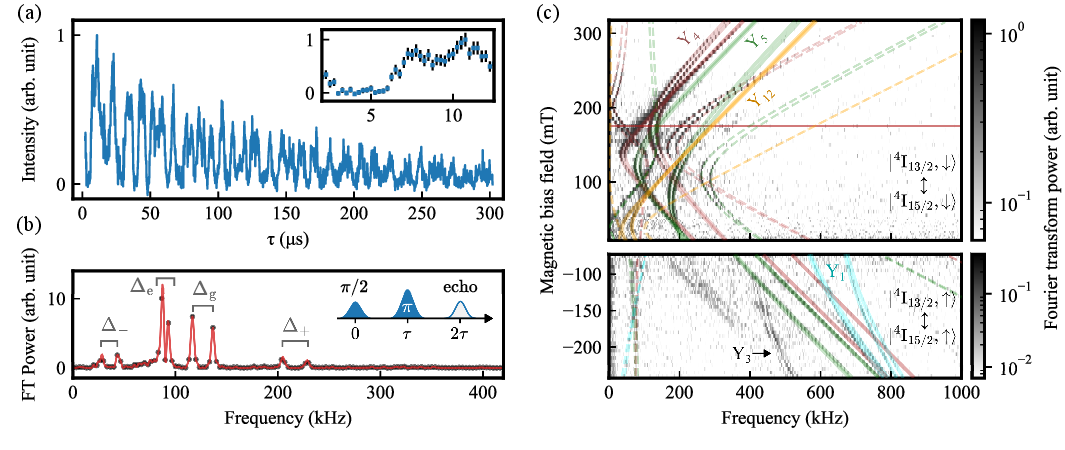}
    \caption{(a) Envelope modulation as observed in the time domain for ensemble measurements at a magnetic bias field of $B=\SI{175}{\milli\tesla}$ applied along the b-axis of YSO. The inset shows a zoom into the first \SI{12}{\micro\second} with error bars corresponding to 1 SD. (b) Fourier spectrum of (a) after subtracting the exponential envelope from the time domain signal without (grey dots) and with (red line) zero-padding to extend the trace length to \SI{1.5}{\milli\second}. All spectral features are attributed to yttrium spin $\mathrm{Y}_4$, which exhibits four frequency components, $\Delta_g$, $\Delta_e$ and $\Delta_\pm:=|\Delta_g \pm \Delta_e|$, each appearing as a doublet due to the splitting of the magnetic classes. (c) Fourier spectrum (with zero-padding) of the envelope modulation as a function of the magnetic field when probing the lower and upper electronic spin transition in the top and bottom panel, respectively.
    Solid lines show fits of the superhyperfine splittings for the yttrium spins $\mathrm{Y}_1$, $\mathrm{Y}_4$, $\mathrm{Y}_5$ and $\mathrm{Y}_{12}$ using Equation~\ref{EQU:hyperbolic} with the fitted parameters shown in Table~\ref{TAB:fitparas}. The dashed lines show the absolute value of sums and differences of the superhyperfine splittings marked in the same color. The horizontal red line marks the position of the spectrum shown in (b).}
    \label{FIG:ensemble_single}
\end{figure*}

To experimentally verify the above reasoning, we start with ensemble measurements. We apply a bias field $\vec{B}^\mathrm{(ext)}$ up to \SI{300}{\milli\tesla} along the b-axis of the crystal. For each magnetic field, we determine the envelope function of the photon echo. For this purpose, we stabilize the excitation laser frequency at the center of the inhomogeneous linewidth of the $\ket{^4\text{I}_{15/2},\downarrow}\leftrightarrow\ket{^4\text{I}_{13/2},\downarrow}$ or $\ket{^4\text{I}_{15/2},\uparrow}\leftrightarrow\ket{^4\text{I}_{13/2},\uparrow}$ transition, where approximately 100 dopants contribute to the photon echo signal. Using an acousto-optical modulator, we generate pulses with a Gaussian intensity envelope with a full-width-at-half-maximum of \SI{0.25}{\micro\second}. The intensity corresponds to $\pi/2-$ and $\pi-$pulses for emitters at the maximum of the optical field in the resonator. We then vary the interpulse delay $\tau$ and record the intensity of the echo. This sequence is repeated 70 (300) times on the lower (upper) electronic spin transition and averaged over all repetitions to obtain a single time trace. An example trace that exhibits a strong envelope modulation and its Fourier transform are shown in Figure~\ref{FIG:ensemble_single}~(a,b). The temporal decay of the signal, and thus the spectral resolution, are limited by the Hahn echo time of the ensemble. Using dynamical decoupling pulses on the optical transition~\cite{ulanowski_spectral_2024}, this could potentially be increased up to the limit set by the optical lifetime $T_2 < 2\cdot T_1$. Figure~\ref{FIG:ensemble_single}~(c) depicts the Fourier spectrum of the envelope modulation as a function of the magnetic field. We observe several very sharp modulation frequencies of a few kilohertz widths. Their values and magnitudes vary smoothly with the applied magnetic field. For each distinct frequency, we find a second close-by line. The reason is a slight deviation of the magnetic field direction from the b-axis of the crystal, with an estimated value of $\lesssim 3\si{\degree}$. This lifts the degeneracy of the two magnetic classes of Er in site 1 in YSO, which results in the observed double-peak features.

To further analyse the spectra of the envelope modulation, we first note that according to Equation~\ref{EQU:Deltas} the superhyperfine splittings $\Delta_\mathrm{g/e}(B)$ with $i \in \{\mathrm{g,e}\}$ follow

\begin{equation}\label{EQU:hyperbolic}
    \Delta_i(B)=\frac{\mu_\mathrm{N}}{h}\left|g_\mathrm{Y}\vec{B}_{i}^\mathrm{(tot)}\right|=\left|g_\mathrm{Y}\right|\frac{\mu_\mathrm{N}}{h}\sqrt{B_{i}^\mathrm{\perp~2}+(B_{i}^\mathrm{\parallel}+B)^2}.
\end{equation}

Here, $B_{i}^\mathrm{\perp}$ and $B_{i}^\mathrm{\parallel}$ are the erbium dipolar field components orthogonal and parallel to the direction of the magnetic bias field $B$, respectively. We can identify several frequency components that follow this hyperbolic relationship, as shown in Figure~\ref{FIG:ensemble_single}~(c). All other strong modulation components correspond to the sums $\Delta_\mathrm{g}(B)+\Delta_\mathrm{e}(B)$ and differences $|\Delta_\mathrm{g}(B)-\Delta_\mathrm{e}(B)|$ of these frequency components. However, the amplitude of the observed peaks significantly deviates from the prediction of Eq. \ref{EQU:modulation} that has been derived for free-space photon echo experiments~\cite{car_optical_2019}.

Still, as expected from the model, the spins $\text{Y}_4$, $\text{Y}_5$ and $\text{Y}_{12}$ with the largest branching contrast are clearly observed when the field is applied along the b-axis and its magnitude is varied. Furthermore, on the upper electronic spin transition we observe weak features that can be unambiguously attributed to the spins $\text{Y}_1$ and $\text{Y}_{3}$ despite their low branching contrast. The reason is that the modulation frequency of these spins differs significantly from that of other Y nuclei, as further discussed in the appendix. 

In Table~\ref{TAB:fitparas}, we compare the values of $B_{i}^\mathrm{\perp}$ and $B_{i}^\mathrm{\parallel}$ extracted from hyperbolic fits with the ones predicted from first principles, finding a good agreement. We can thus correctly identify the individual Y nuclear spins. For $\text{Y}_{3}$, the overlap of the frequency components of ground- and excited states precludes a reliable fit.
At large magnetic fields ($B \gg -B_{i}^\mathrm{\parallel}$), we find that the slope of the curves matches earlier measurements of the gyromagnetic ratio of Y nuclear spins, which slightly differs from the value of an isolated nucleus because of the local magnetic field shielding of the surrounding electrons~\cite{muller_nuclear_2021} (see appendix).

However, at small fields we observe a slight deviation of the superhyperfine coupling that cannot be fully explained by the directional misalignment of the magnetic bias field. 
This may indicate that the interaction of the electronic spins with the nearest neighbors in YSO slightly deviates from that of purely dipolar couplings. Quantitative modeling, however, would require additional measurements at different magnetic field orientations.

\begin{table}
    \centering
    \begin{tabular}{@{}clS[table-format=3.0]@{\hspace{9pt}}S[table-format=3.0]@{\hspace{9pt}}S[table-format=3.0]@{\hspace{9pt}}S[table-format=3.0]@{\hspace{9pt}}@{}}
    \toprule
    {Spin} &        & {$B_\mathrm{g}^\mathrm{\parallel}$ (mT)} & {$B_\mathrm{g}^\mathrm{\perp}$ (mT)} & {$B_\mathrm{e}^\mathrm{\parallel}$ (mT)} & {$B_\mathrm{e}^\mathrm{\perp}$ (mT)} \\ \midrule
    
    \multirow{3}{*}{$\mathrm{Y}_{1}$} 	 & pred. 
	 &  -5 & 317 & 35 & 250 \\ 
	 & obs. I &  -11  & 315 & 36 & 248 \\ 
	 & obs. II &  4  & 314 & 44 & 248 \\ 
	\midrule
    \multirow{3}{*}{$\mathrm{Y}_{4}$} 	 & pred. 
	 &  164 & 49 & 131 & 23 \\ 
	 & obs. I &  169  & 55 & 135 & 17 \\ 
	 & obs. II &  166  & 65 & 135 & 22 \\ 
	\midrule
    \multirow{3}{*}{$\mathrm{Y}_{5}$} 	 & pred. 
	 &  128 & 52 & 83 & 83 \\ 
	 & obs. I &  122  & 36 & 78 & 82 \\ 
	 & obs. II &  121  & 42 & 72 & 87 \\ 
	\midrule
    \multirow{3}{*}{$\mathrm{Y}_{12}$} 	 & pred. 
	 &  41 & 31 & 35 & 17 \\ 
	 & obs. I &  40  & 31 & 36 & 16 \\ 
	 & obs. II &  40  & 31 & 36 & 18 \\ 
	\bottomrule

    \end{tabular}
    \caption{Predicted and observed values of $B_\mathrm{i}^\mathrm{\perp}$ and $B_\mathrm{i}^\mathrm{\parallel}$ as defined in Equation~\ref{EQU:hyperbolic}. There are two observed parameters for each of the predicted ones, as for the nominal bias field direction there is just a single magnetic magnetic class of erbium dopants, but two slightly non-degenerate classes are observed experimentally because of a slight field misalignment.}
    \label{TAB:fitparas}
\end{table}

\section{OEEM on single emitters}
The coupling to a high-finesse cavity not only allows for ensemble measurements with very low dopant concentrations but even for probing the nuclear spin environment of individual emitters. To this end, we tune the laser to the edge of the inhomogeneous linewidth, where individual dopants are spectrally well-isolated and can thus be addressed individually \cite{ulanowski_spectral_2022}. Following our previous work~\cite{ulanowski_spectral_2024}, we investigate classes of emitters that are strain-shifted away from center of the inhomogeneous line by co-doping with 100~ppm of europium. This leads to distinct satellite lines with a tailored concentration of emitters while retaining the excellent optical coherence properties of undoped samples~\cite{merkel_coherent_2020}.

\begin{figure}
    \centering
    \includegraphics[width=1\columnwidth]{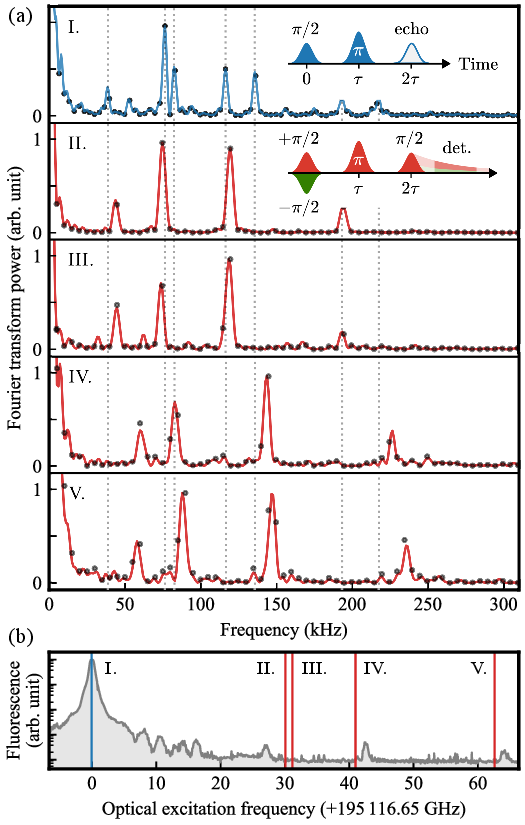}
    \caption{(a) OEEM spectra for several individual spins (red) compared to ensemble measurements (blue) at the center of the inhomogeneous line for a magnetic bias field of \SI{170}{\milli\tesla} applied along the b-axis of YSO. The Fourier transform of the time domain data is shown without (gray dots) and with (solid lines) zero-padding with a ratio of 1 to 8. (b) Optical transition frequency of the probed single emitters (red lines) relative to the center of the inhomogeneous distribution (grey) where the ensemble measurement was performed (blue line, from~\cite{ulanowski_spectral_2024}). In panel (a), a clear shift of the OEEM modulation frequencies is observed that is more pronounced for dopants with a larger optical frequency shift.}
    \label{FIG:singleions}
\end{figure}

Thus, we continue with OEEM on individual emitters in the satellite lines. As two-pulse echo sequences cannot be used with single emitters, we switch to fluorescence-detected Hahn echos~\cite{montjovet-basset_incoherent_2024}, as in our earlier single-emitter experiments~\cite{ulanowski_spectral_2022, ulanowski_spectral_2024}. The technique differs from our earlier measurements by a second $\pi/2$ pulse that is applied at the end of the Hahn echo sequence at the point of rephasing, mapping the coherence in the system onto a population that we detect as cavity-enhanced spontaneous emission. By subtracting the amount of fluorescence observed for a relative phase of $\pi$ between the first and last $\pi/2-$pulse from the fluorescence when the two $\pi/2-$pulses share the same phase, we yield a quantity that is proportional to the expectation value of the Pauli operator $\hat{\sigma}_x$ at the point of rephasing. As such, this technique allows us to record a photon-echo-like signal even for a single emitter, and characterize the superhyperfine interaction via its envelope modulation.

We performed this measurement on many individual emitters at a magnetic bias field of \SI{170}{\milli\tesla}, where a single Y dominates the envelope modulation. The time traces can be accessed in the data repository associated with the manuscript. Figure~\ref{FIG:singleions}~(a) shows the ensemble data (blue, panel I) and four representative single-emitter spectra (red, II-V) that were recorded at different detunings from the center of the inhomogeneous distribution shown in panel (b). The single-dopant measurements shows half as many distinct peaks, as any emitter only belongs to a single magnetic class, while both classes contribute to the ensemble spectra. 

While the frequency of the single-emitter peaks at moderate detuning from the center of the inhomogeneous line, panel II and III, match the ensemble measurement (grey dashed) quite well, a significant shift is observed at larger detunings, panels IV and V. Apparently, the Er-Y interaction is modified in the proximity of an additional Eu dopant. This can be caused by two effects: First, a local distortion of the crystal structure which modifies the relative position of Er and its surrounding Y neighbors; Second, a modification of the crystal field Hamiltonian and thus the magnetic moment of the respective erbium dopant. The latter effect has been previously observed as a slightly modified Zeeman interaction for erbium dopants belonging to satellite lines~\cite{ulanowski_spectral_2024}. Still, further measurements would be required to exclude a significant contribution of the first effect.

\section{Discussion}

In summary, we have demonstrated OEEM with an unprecedented signal-to-noise ratio and a spectral resolution down to a few kHz, which is enabled by embedding an Er:YSO crystal in a high-finesse cavity. At the level of individual emitters, we observed that superhyperfine spectra in satellite peaks induced by close-by europium co-dopants can differ significantly from those obtained at the center of the inhomogeneous line. In the future, our technique can give further unprecedented insights into the interaction of optically addressable spins with their magnetic environment. As an example, it may facilitate detailed investigations of contact interactions or atomic position shifts induced by single impurities with {\AA}ngstrom precision. To this end, the spectroscopic signal would need to be further improved by adding a vector magnet, which would facilitate magnetic field sweeps and optimization in arbitrary directions. In addition, this would allow further tailoring of the branching contrast, such that selected nuclear spins can be used as a resource in quantum technology, enabling, e.g., the entanglement of nuclear spins with emitted photons. In particular, this technique may be applied to target the Eu co-dopants in our sample that may then serve as quantum memory with exceptional coherence \cite{zhong_optically_2015}.

While our study has been performed on the specific platform of Er:YSO, the presented techniques can be directly transferred to a wide variety of solid-state systems. When using the simple two-pulse envelope modulation technique implemented above, with other emitters or hosts the spectral resolution may be limited by the optical coherence. This limitation can be overcome by transferring the three-pulse sequences known from ESEEM \cite{dikanov_electron_2024} to the optical domain following our approach. Thus, by eliminating the restrictions on the magnetic fields imposed by the ESEEM technique and simultaneously enabling single-emitter measurements, we have demonstrated that OEEM allows for detailed and microscopic studies of nuclear spin baths and, thus, decoherence dynamics of spins in solids.

\section{Acknowledgements}
We acknowledge discussions with Oliver Diekmann, Stefan Rotter, and Thierry Chaneli\`ere. This project received funding from the Deutsche Forschungsgemeinschaft (DFG, German Research Foundation) under the German Universities Excellence Initiative - EXC-2111 - 390814868 and via grant agreement 547245129, from the European Research Council (ERC) under the European Union's Horizon 2020 research and innovation programme (grant agreement No 757772) and from the Munich Quantum Valley, which is supported by the Bavarian state government with funds from the Hightech Agenda Bayern Plus.

\section{Data availability}
The datasets generated and analyzed during the current study are openly available in the mediaTUM repository \cite{mediatum_data}.

\appendix*
\section{First-principle calculation of the OEEM signals}
\label{SEC:appendix}

In the following, we calculate the expected spectral features in the OEEM signal for each of the fifteen closest yttrium spins around a single erbium dopant, which enables assigning the measured spectral features in Figure \ref{FIG:ensemble_single}~(c) to individual Y spins. To this end, we use the known positions of the Y sites given in Table \ref{TAB:yttrium} to calculate the dipolar field induced by the Er electron spin at the location of each nuclear spin individually. This field is then split into the components $B_{\mathrm{g/e}}^\mathrm{\perp}$ and $B_{\mathrm{g/e}}^\mathrm{\parallel}$, which are orthogonal and parallel to the externally applied bias field. We assume $\vec{B}^\mathrm{(ext)}\parallel\mathrm{b}$, so the degeneracy of the two magnetic classes is not lifted in the model. Using Equation \ref{EQU:hyperbolic} we then determine the superhyperfine splittings,  $\Delta_\mathrm{g}(B)$ and $\Delta_\mathrm{e}(B)$, for the ground and optically excited state, respectively. 

\begin{figure*}
    \centering
    \includegraphics[width=1\textwidth]{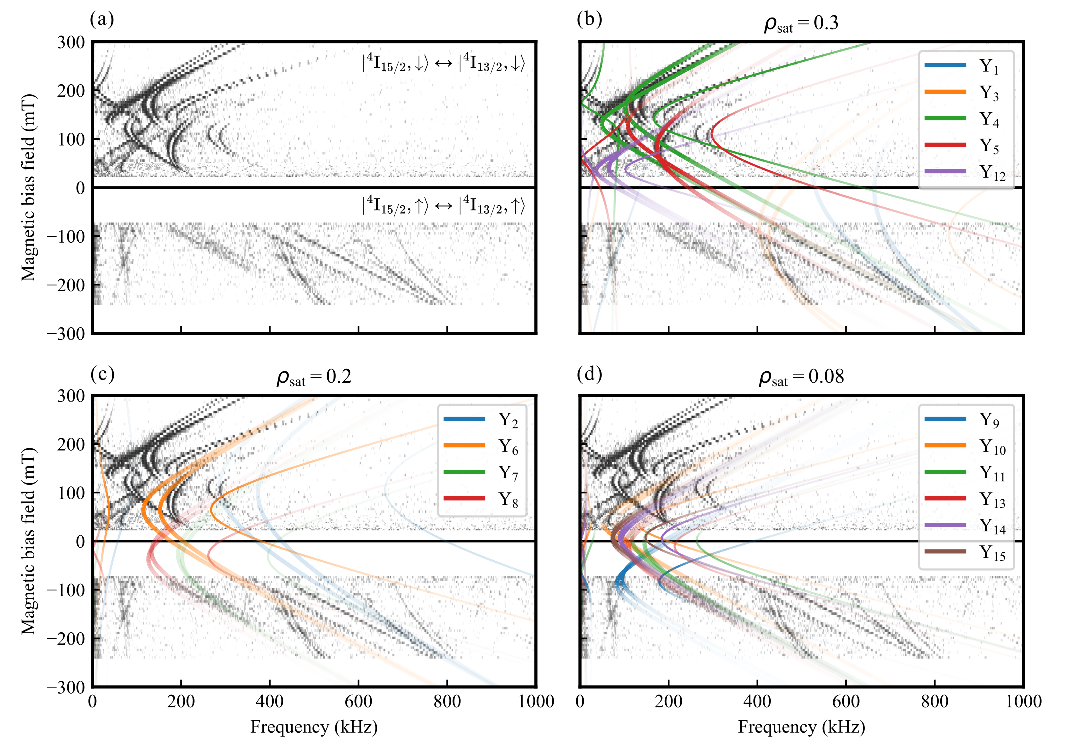}
    \caption{(a) Fourier spectra of the measured envelope modulation as a function of the magnetic field magnitude for both electronic Er spin transitions. The presented data and color scale are identical to Figure \ref{FIG:ensemble_single}~(c) but are shown without annotations for clarity. (b-d) Measured Fourier spectra and predicted frequency components of individual Y spins when the magnetic field is applied exactly along the b-axis of the crystal. Bold solid lines indicate the level splittings $\Delta_\mathrm{g/e}$ and thin solid lines their sum and difference: $|\Delta_\mathrm{g} \pm \Delta_\mathrm{e}|$. The color intensity of the lines represents the branching contrast $\rho$ of the respective Y spin. For $\rho=0$, the line is fully transparent, and its intensity increases linearly with the branching contrast until it reaches full saturation for $\rho = \rho_{sat}$. The latter is indicated above the panels.}
    \label{FIG:ensemble_model}
\end{figure*}

These splittings and their sum and difference, $|\Delta_\mathrm{g}(B)\pm\Delta_\mathrm{e}(B)|$, are shown as an overlay to the measured data in Figure \ref{FIG:ensemble_model}. Four separate panels improve the readability. The individual spins are represented by colored lines and the field-dependence of the branching contrast is visualized by the color intensity. Figure \ref{FIG:ensemble_model}~(a) shows the bare data; panel (b) includes the most prominent spins, $\text{Y}_4$, $\text{Y}_5$ and $\text{Y}_{12}$, and the well-separated spins with smaller branching contrast, $\text{Y}_1$ and $\text{Y}_3$. Figure \ref{FIG:ensemble_model}~(c) and (d) show the remaining yttrium spins up to $\text{Y}_{15}$. To increase the visibility, we use a different intensity scaling of the solid lines, as indicated above the respective panels.

Notably, we do not observe the $\Delta_\mathrm{g/e}$ components of spin $\text{Y}_6$ which is expected to exhibit a branching contrast up to 0.2. Only a weak signature of the difference component can be found on the lower electronic spin branch. The other components are in close proximity to, and may thus overlap with, strong features that originate from other strongly coupled spins. In addition, a group of spins (including $\text{Y}_7$, $\text{Y}_8$, $\text{Y}_{10}$, $\text{Y}_{11}$ and $\text{Y}_{13}$-$\text{Y}_{15}$) has features in a similar spectral region. Thus, the "band-like" feature observed between $\SI{200}{\kilo\hertz}$ at approximately $\SI{-100}{\milli\tesla}$ and $\SI{400}{\kilo\hertz}$ at $\SI{-200}{\milli\tesla}$ cannot be assigned to individual nuclei. The remaining spins $\text{Y}_2$ and $\text{Y}_9$ are spectrally well-separated from the spins assigned in Figure \ref{FIG:ensemble_model}~(b); however, they do not exhibit a clear signature and would require a better signal-to-noise ratio to be observed. 

For Y spins that are even further away from the Er dopant, the
$B_{\mathrm{g/e}}^\mathrm{\perp}$ and $B_{\mathrm{g/e}}^\mathrm{\parallel}$ components get so small that a significant branching contrast is only expected at close-to-zero magnetic fields. However, this regime is not accessible in our experiment for two reasons: First, the large number of coupled spins results in a superhyperfine-induced collapse of the photon echo \cite{car_superhyperfine_2020}. Second, the reduced phononic relaxation via the direct process \cite{wolfowicz_quantum_2021} would lead to long electron spin lifetimes. Thus, spectral hole burning effects would decrease the signal-to-noise ratio in this regime.

A contribution of $^{29}\mathrm{Si}$ nuclear spins to the observed spectral features can be excluded due to their low abundance ($<$5~\%). Furthermore, the 4-fold larger nuclear g-factor of $^{29}\mathrm{Si}$ compared to $^{89}\mathrm{Y}$ would imply a higher magnetic field sensitivity which is not observed in the spectra.

\begin{figure}
    \centering
    \includegraphics[width=1\columnwidth]{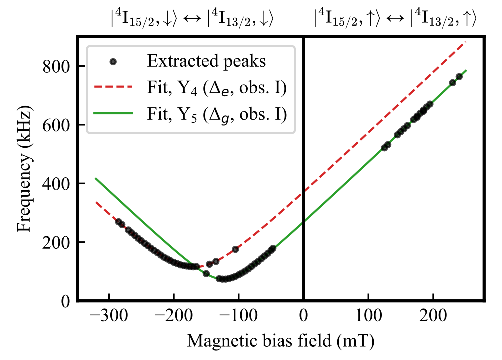}
    \caption{Measurement of the gyromagnetic ratio of $^{89}$Y. Based on equation \ref{EQU:hyperbolic}, a hyperbolic fit of two different Y spin spectra (indicated in the legend), yields \SI{-2.090(6)}{\mega\hertz\per\tesla} for Y$_4$ (dashed red line) and \SI{-2.087(2)}{\mega\hertz\per\tesla} for Y$_5$ (solid green line).}
    \label{FIG:gY_fit}
\end{figure}

\section{$^{89}\mathrm{Y}$ gyromagnetic ratio in YSO}
\label{SEC:appendix_gY}

The gyromagnetic ratio of $^{89}\mathrm{Y}$ in YSO is determined using a fit based on equation \ref{EQU:hyperbolic}, where $B_{i}^\mathrm{\perp}$, $B_{i}^\mathrm{\parallel}$ and $g_\text{Y}$ are free parameters. We compare two different Y spins of class I, $\mathrm{Y}_4$ and $\mathrm{Y}_5$, which exhibit a high signal-to-noise ratio in the measured spectra in Figure \ref{FIG:ensemble_single}~(c). The fit in Figure \ref{FIG:gY_fit} gives an average gyromagnetic ratio of $\SI{-2.089(3)}{\mega\hertz\per\tesla}$, where a negative sign is assigned based on earlier measurements \cite{hasler_89yttrium_1977}.

Within errors, this agrees with the expectation for the two crystallographic sites in YSO, $\SI{-2.0868}{\mega\hertz\per\tesla}$ and $\SI{-2.0866}{\mega\hertz\per\tesla}$. Here, we use chemical shifts of 237 ppm and 150 ppm, which are known from nuclear magnetic resonance measurements of the X2-Y$_2$SiO$_5$ polymorph \cite{becerro_revisiting_2004} and referenced to a solution of YCl$_3$. In the latter, $^{89}\mathrm{Y}^{3+}$ exhibits a gyromagnetic ratio of $\SI{-2.0863}{\mega\hertz\per\tesla}$ \cite{hasler_89yttrium_1977}. We note that the difference to the value of an isolated nucleus, $\SI{-2.0949}{\mega\hertz\per\tesla}$ \cite{raghavan_table_1989}, can be largely attributed to the shielding of the magnetic field by the electronic orbitals, with a shielding factor of 1.0041 for $^{89}\mathrm{Y}$.

\newpage

\bibliographystyle{apsrev4-2}
\bibliography{bibliography.bib}

\end{document}